\documentclass{article}
\usepackage{spconf,amsmath,graphicx,array,booktabs,makecell}

\usepackage[hyphens]{url}
\usepackage[hidelinks]{hyperref}
\hypersetup{breaklinks=true}

\title{SPEECH ARTIFACT REMOVAL FROM EEG RECORDINGS OF SPOKEN WORD PRODUCTION WITH TENSOR DECOMPOSITION}

\name{Holy Lovenia$^{1, 3}$, Hiroki Tanaka$^{1}$, Sakriani Sakti$^{1, 2}$, Ayu Purwarianti$^{3}$, and Satoshi Nakamura$^{1, 2}$}
\address{$^{1}$ Nara Institute of Science and Technology, Japan \\ $^{2}$ RIKEN, Center for Advanced Intelligence Project AIP, Japan \\ $^{3}$ Department of Informatics, Bandung Institute of Technology, Indonesia \\ \texttt{\small{holy.lovenia@gmail.com, \{hiroki-tan, ssakti, s-nakamura\}@is.naist.jp, ayu@stei.itb.ac.id}}}

\begin{document}

\maketitle

\begin{abstract}
Research about brain activities involving spoken word production is considerably underdeveloped because of the undiscovered characteristics of speech artifacts, which contaminate electroencephalogram (EEG) signals and prevent the inspection of the underlying cognitive processes. To fuel further EEG research with speech production, a method using three-mode tensor decomposition (time $x$ space $x$ frequency) is proposed to perform speech artifact removal. Tensor decomposition enables simultaneous inspection of multiple modes, which suits the multi-way nature of EEG data. In a picture-naming task, we collected raw data with speech artifacts by placing two electrodes near the mouth to record lip EMG. Based on our evaluation, which calculated the correlation values between grand-averaged speech artifacts and the lip EMG, tensor decomposition outperformed the former methods that were based on independent component analysis (ICA) and blind source separation (BSS), both in detecting speech artifact (0.985) and producing clean data (0.101). Our proposed method correctly preserved the components unrelated to speech, which was validated by computing the correlation value between the grand-averaged raw data without EOG and cleaned data before the speech onset (0.92-0.94).
\end{abstract}


\section{Introduction}
\label{sec:intro}

Many cognitive processes are involved in the act of speaking, such as phonetic encoding and articulation \cite{2}, even when the utterances are simply naming a common object, e.g., dog \cite{1}. To uncover the representations and brain processes underlying word production in speech, the inspection of cognitive processes in word production can be supported by tools, including an EEG. An EEG can obtain a high temporal resolution \cite{3}, which enables research of brain activity with a precise time course. Unfortunately, research in EEG-related fields involving such overt speech as language and word production suffers from noise caused by non-brain activities. EEG experiments always encounter problems because of its low signal-to-noise ratio (SNR). Even without speaking, only eye blink, the SNR is already low.

Moreover, many muscle movements are required to produce speech, e.g., mouth, jaw, and tongue movements \cite{3}. As a result, the cognitive components of spoken word production processes are inevitably contaminated by these muscle movements \cite{1}, which will be addressed as speech artifacts below. Overlapped by speech artifacts, it is difficult to tell whether the components from the recorded data are real brain activities or merely noise. To avoid the occurrence of speech artifacts, many works on EEG data have experimented with covert speech planning and delayed speaking tasks. Over the years, covert speech experiments have undeniably sparked essential discoveries by EEG studies. However, tasks using immediate overt speech are preferred in several situations. In overt speech production, no altering or omitting of the cognitive steps (e.g. articulation) involved in speech production is needed because speech is produced naturally. Hence, speech artifact removal methods must free EEG data from speech-related artifacts during overt speech production tasks. Vos et al. utilized BSS and canonical correlation analysis to separate the muscle movements regarding speech from the raw data \cite{6}. A more recent study by Porcaro et al. exploited ICA to attenuate speech artifacts and investigated the original sources that are related to the act of speaking \cite{1}. Both methods are based on matrix decomposition \cite{1}\cite{6}.

EEG recordings normally have at least two modes, time and space. The channel (or component) is represented by the space mode, and the brain activity across time is described by the time mode. In matrix decomposition methods, such as BSS and ICA used in previous works, both of these modes extract the origin sources of brain activity. However, besides time and space mode, there can be more modes involved in EEG experiments, such as frequency, condition, group, and subject. Most signal-processing methods only support two-way data instead of three or more dimensions. Thus, efforts have tried to fit all of the required modes into the two-dimensional EEG data. Various attempts have unfolded a multi-way array into a matrix, including concatenation and stacking to the space/time mode \cite{4}. A tensor is a multi-way array that is regarded as high-order when it has more than two modes. EEG observation of the temporal, spectral, spatial, and other changes in brain activities can be done simultaneously with a tensor \cite{4}. Tensor factorization also enables the consideration of more than two modes during decomposition \cite{maki}. The present study has two main objectives. The first is to propose a new method for removing speech artifacts. The proposed method is based on tensor decomposition for sources reconstruction. The second is to evaluate the performance of the proposed method (using tensor decomposition) with the existing methods (using ICA and BSS). Then we validated the cleaned data obtained by all of the methods to ensure the preservation of brain signals.

\section{PREVIOUS METHODS TO REMOVE SPEECH ARTIFACTS}
\label{sec:previous-methods}

The baselines used in this research came from an adapted method of SAR-ICA by Porcaro et al. \cite{1} and the automatic version of BSS-CCA by Vos et al. \cite{6}.

\subsection{Speech artifact removal by independent component analysis (SAR-ICA)}
\label{ssec:sar-ica}

SAR-ICA, which is a speech artifact removal (SAR) method that uses ICA for decomposing brain sources \cite{1}, is comprised of four main phases: decomposition, artifact detection, control cycle, and clustering. The sources reconstruction used fastICA, and the corresponding independent components (ICs) were marked as artifacts based on their statistical and spectral characteristics, with the electro-oculogram (EOG) and electromyogram (EMG) channels as the benchmarks. In addition, we also visually inspected the averaged trials, the single trials, and the topographical distributions of the ICs to manually categorize whether the corresponding IC was an artifact. After the identification of both the eye-related (ocular) and speech-related artifacts, we applied a control cycle. This step confirmed that the differences between the raw and the artifact-free data were only caused by the artifact removal. If any brain activity was present in the removed data, we reduced the thresholds used for the artifact identification and repeated the corresponding step. Then all of the components were clustered based on their result from the previous steps. After the cluster membership, each cluster was backprojected to the channel space.

\subsection{Blind source separation with canonical correlation analysis (BSS-CCA)}
\label{ssec:bss-cca}

As input data used for the BSS-CCA method, we used Cluster 3 (raw data without EOG) by SAR-ICA, which consisted of raw data without ocular artifacts. As its name implies, this method based its decomposition process on BSS \cite{6}. Instead of assuming statistical independence, BSS-CCA uses a different basis to separate the brain sources: autocorrelation properties by canonical correlation analysis (CCA). This method assumes that EMG activity is weakly autocorrelated over time, and brain activity is more likely to be autocorrelated because of its coherence over time. There is another criterion to select the components related to muscle movements. If the average power of the components in the EMG band (approximated by 15-30 Hz) equals or exceeds 1/n (default n is 7) of the average power in the EEG band (approximated by 0-15 Hz), the corresponding component is marked as a muscle (speech-related) artifact. By defining a quantitative criterion to mark the speech artifacts, no visual inspection was needed in BSS-CCA.

\section{PROPOSED METHOD USING TENSOR DECOMPOSITION}
\label{sec:proposed-method}

Unlike the previous methods that used matrix decomposition, the proposed method carried out the sources reconstruction step using tensor decomposition, which is also generated from blind source separation \cite{4}. However, the concept behind these two methods is essentially different. A brief overview of the differences between the baselines and the proposed method is represented in Table \ref{tab:comparison}. In the proposed method, we applied third-order canonical polyadic (CP) decomposition from the Tensor Toolbox to the original data \cite{7}\cite{8}.

\begin{table}[h]
\caption{Overview comparison of SAR-ICA \cite{1}, BSS-CCA \cite{6}, and proposed method with tensor decomposition (TD)}
\begin{tabular}{l l l l}
\toprule
\thead{} & \thead{SAR-ICA} & \thead{BSS-CCA} & \thead{TD} \\
\midrule
Decomposition method & ICA & BSS & CPD \\
\midrule
EMG channels & Yes & No & Yes \\
\midrule
Needs visual inspection & Yes & No & Yes \\
\bottomrule
\end{tabular}
\label{tab:comparison}
\end{table}

\begin{figure}[ht]
\centering
\includegraphics[width=\linewidth]{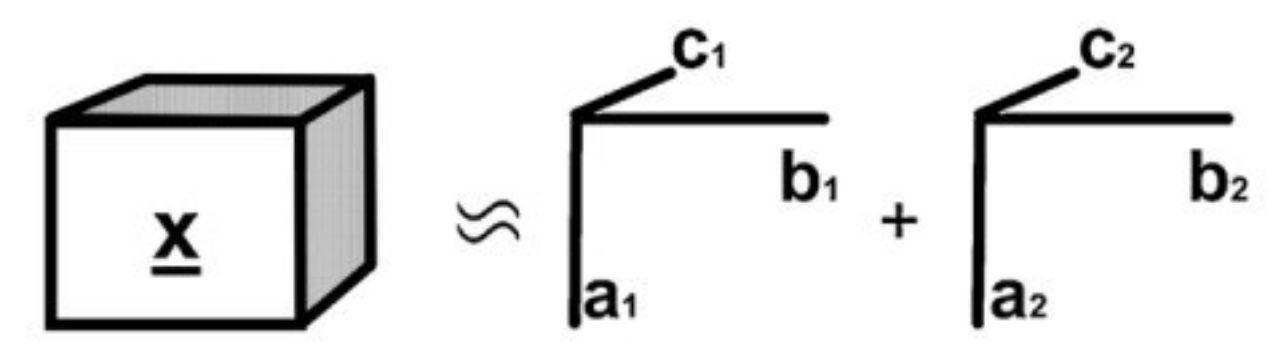}
\caption{Third-order tensor with two CPD components}
\label{fig:cpd}
\end{figure}

In CP decomposition (CPD), which is also called canonical decomposition (CANDECOMP) or parallel factor analysis (PARAFAC) \cite{4}, the number of components for each mode is the same \cite{4}\cite{9} (Fig. \ref{fig:cpd}). When the original data are decomposed by CPD with time, frequency, and channel modes, then the number of temporal, spectral, and space components is identical. Given $N$th-order tensor $X$, CPD is generally described as
\begin{equation} \label{eq:1}
\begin{split}
X & = \sum_{r=1}^{R} u_r^{(1)} \bullet u_r^{(2)} \bullet \cdots \bullet u_r^{(N)} + E = \sum_{r=1}^{R} X_r + E \\ & = \hat{X} + E \approx \hat{X},
\end{split}
\end{equation}
where $r$ is the number of components, $n = 1, 2, \cdots, N$ is the mode, and $E$ is the residual \cite{4}.

The number of extracted components for each subject was determined by the difference of the fit (DIFFIT) method by Timmerman et al. with a lower bound of 5 and an upper bound of 26 \cite{4}\cite{10}. We chose five to prevent the number of components from being too low and 26 EEG channels. The number of extracted components used by the matrix decomposition was also beneficial when applying CP decomposition, such as ICA, which uses the number of sources in the brain \cite{4}. The average number of extracted components from the EEG data across the subjects was approximately 15. The lowest estimation was 8, and the highest number of extracted components was 24. The artifact detection was based on three aspects: the frequency domain characteristics, the time domain characteristics, and a visual inspection. The frequency domain characteristics were observed by calculating the correlation coefficients between the spectral components of the EEG data and the spectral component of the EMG data.

Similar observations analyzed the time domain characteristics. However, there was a significant differentiator: the usage of speech onset. The Pearson correlation values between the temporal components of the EEG data and the temporal component of the EMG data were calculated starting from the averaged speech onset of the corresponding subject until the next second or the end of the trial, whichever happened first. If the correlation coefficient showed a strong association between both the temporal and spectral components and the EMG data, the corresponding Kruskal tensor (the CPD decomposition result of the original tensor \cite{7}) was marked as a speech artifact. We manually inspected the topographical distributions in the space mode and averaged the trial components. Then we based the decision whether to mark the Kruskal tensor as a speech artifact on all of the results.

The difference between before and after removing the Kruskal tensors was marked as speech artifact and observed as a post-removal validation. If a brain signal was caught in the removed Kruskal tensor, then the artifact detection step was repeated. After the previous steps were done, the Kruskal tensors were classified into three different clusters. Cluster 1 (raw data without ocular artifact) contained the average of all the Kruskal tensors. Cluster 2 (speech artifact) was composed of the Kruskal tensors marked as muscle-contaminated. Cluster 3 (cleaned data) consisted of artifact-free Kruskal tensors.

\section{EXPERIMENT}
\label{sec:exp-design}

\subsection{Participants and materials}
\label{ssec:exp-participants-materials}

Nine native Japanese speakers (mean age 23.3, SD 2.6, 8 males, 2 females) with normal or corrected-to-normal vision participated in the experiment. Eight speakers were right-handed, and the rest were corrected-to-right. One subject was removed from further analysis due to a high rate of errors. We used 45 different line drawings of common objects as the stimuli \cite{5}.

\subsection{Picture-naming task}
\label{ssec:exp-flow}

The raw data collection consisted of pre-experiment and experiment phases. In the pre-experiment phase, the subjects looked at a booklet of picture-name pairs and task description. This phase confirmed that the participants knew the Japanese names of the picture stimuli used in the experiment. The subjects were instructed to only blink after naming the displayed stimulus and to avoid body and head movements as much as possible. There were two blocks in the experiment phase: practice and experiment.

Each block was composed of 45 trials. A rest period was provided between the blocks. In this period, the subjects rested their eyes and body before moving to the next part of the experiment. Each trial randomly used one of the line drawings as the picture stimulus. Throughout the block, each picture appeared once in a random order. At the beginning of every trial, a fixation cross appeared for one second. Then it was replaced with a picture stimulus. The participant named the displayed stimulus as quickly and as accurately as possible. The stimulus remained on display for three seconds. The stimulus presentation was coded using Presentation (Neurobehavioral Systems).

\subsection{EEG, EMG, and EOG recordings}
\label{ssec:exp-recording-tools}

The EEG data were recorded with ActiCAP as the EEG cap and BrainAmp DC (BrainProducts) as the amplifier. During our experiment, 27 electrodes were used as EEG channels on the scalp and one as the default reference. Two electrodes monitored the electro-oculogram (EOG), also known as eye artifact (such as eye blinks), just above and below the left eye. A lip electromyogram, which is a speech artifact (muscle movements caused by speech), was recorded by placing two electrodes at the left orbicularis oris superior (OOS) and the left orbicularis oris inferior (OOI) halfway between the center and the corner of the mouth.

\section{DATA ANALYSIS}
\label{sec:prod-results}

The data analysis was done using the FieldTrip \cite{Oostenveld2011FieldTripOS} and Tensor Toolbox \cite{7}.

\subsection{Data pre-processing}
\label{ssec:preproc}

The continuous raw data collected from the picture-naming experiment were epoched into trials that began at -1 second and ended at 3 seconds. We removed from further analysis trials that contained incorrect names, missed answers, or self-repair attempts and performed baseline correction and 0.1-30 Hz bandpass filtering. EEG channels were offline re-referenced to the average of the left and right mastoids, and the EOG and EMG data used bipolar derivation as a reference. All of the data were offline resampled to 512 Hz.

\subsection{Evaluation}
\label{ssec:eval}

We compared the performances of the methods by calculating the Pearson correlation (R) between the grand-average clusters (speech artifact and the cleaned data) and the lip EMG for 0-1350 ms in the time domain. We assumed that the lip EMG represented the real speech artifact. The clusters were averaged across the subjects. The aim of further validation is to confirm whether the cleaned data cluster preserved the brain signals. Validation computed the correlation values between Cluster 1 (raw data without ocular artifacts) and Cluster 3 (cleaned data) by the proposed method for 0-700 ms (before the earliest speech onset) and 0-900 ms (before the grand-average of the speech onsets). If no brain signal was removed in the cleaned data, they are supposedly highly correlated.

\section{RESULTS AND DISCUSSION}
\label{sec:results-discussion}

The table below presents the detailed results of the evaluation step, which is the absolute Pearson correlation value between the speech artifacts and the cleaned data with the lip EMG in the time domain for 0-1350 ms. A higher correlation value is better for the speech artifact, because it shows an association with the lip EMG. For the cleaned data, a low correlation value is better. Proposed method is written as CPD in bold.

\begin{table}[h]
\caption{Grand-average correlation to lip EMG (all, p $<$ 0.01)}
\centering
\begin{tabular}{l l}
\hline
\toprule
\thead{Grand-average data} & \thead{R (0-1350 ms)} \\
\midrule
SAR-ICA's Cluster 2 (Speech Artifacts) & 0.875 \\
\textbf{CPD's Cluster 2 (Speech Artifacts)} & \textbf{0.985} \\
\midrule
SAR-ICA's Cleaned Data & 0.351 \\
BSS-CCA's Cleaned Data & 0.413 \\
\textbf{CPD's Cleaned Data} & \textbf{0.101} \\
\bottomrule
\end{tabular}
\label{tab:evaluation}
\end{table}

\begin{figure}[h]
\centering
\includegraphics[width=\linewidth]{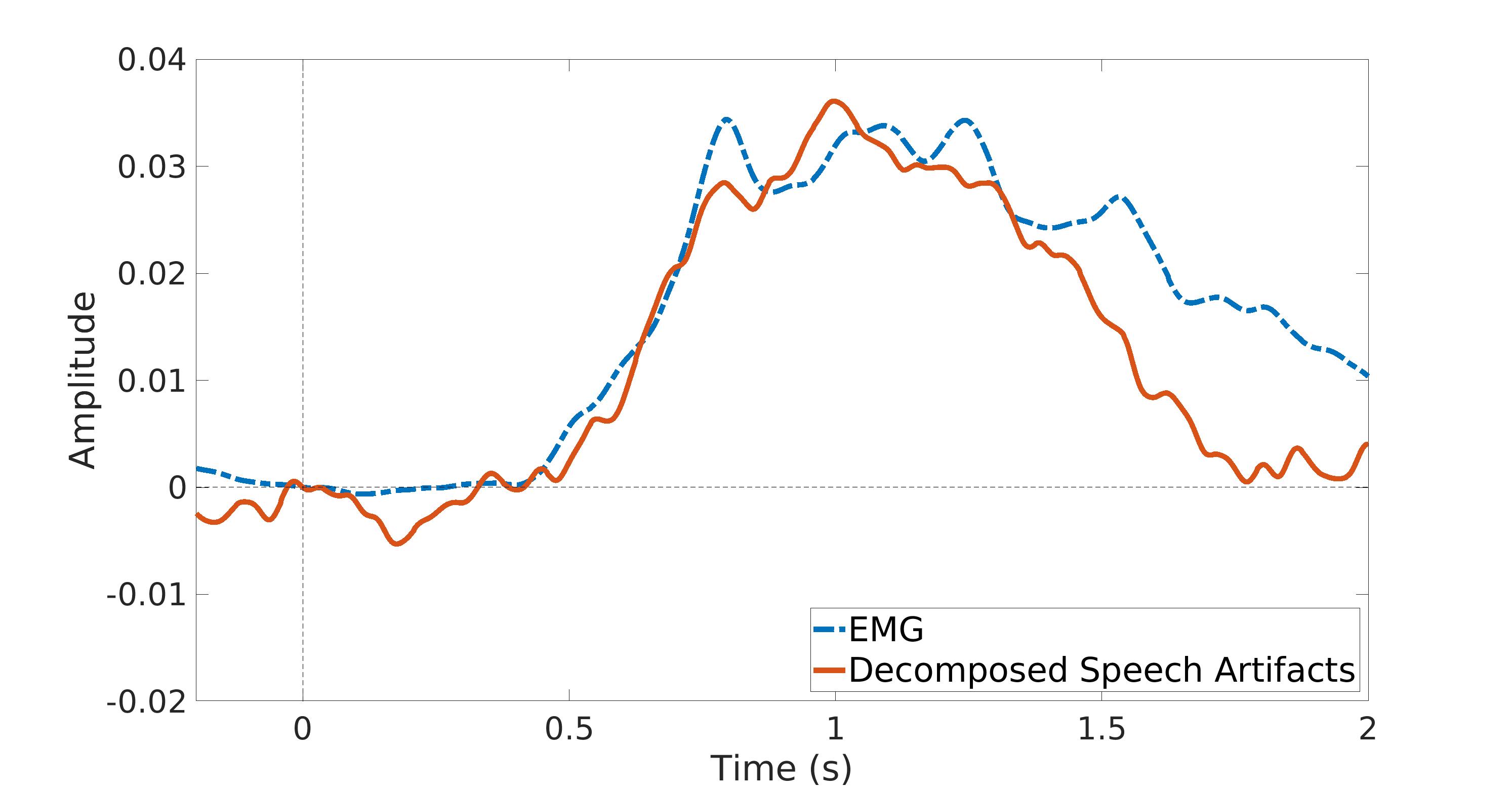}
\caption{Comparison of normalized grand-average lip EMG and decomposed speech artifacts by proposed method}
\label{fig:compare-emg-speech}
\end{figure}

As shown in Table \ref{tab:evaluation}, the proposed method using tensor decomposition outperformed its baselines, SAR-ICA and BSS-CCA, both in detecting speech artifacts (0.985) and producing cleaned data (0.101). Fig. \ref{fig:compare-emg-speech} shows that at most of the time points, the speech artifacts were almost identical to the lip EMG. However, only a slight difference remains between them, which might be caused by the inability of EMG channels to pick up every single movement related to speech.

\begin{table}[h]
\caption{Grand-average correlation to raw data without EOG before speech onset (p $<$ 0.01)}
\centering
\begin{tabular}{l l l}
\hline
\toprule
\thead{Grand-average data} & \thead{R (0-700 ms)} & \thead{R (0-900 ms)} \\
\midrule
CPD's Cleaned Data & 0.927 & 0.942 \\
\bottomrule
\end{tabular}
\label{tab:validation}
\end{table}

This difference was also expressed in the validation result in Table \ref{tab:validation}, which yielded a correlation value of approximately 0.92-0.94 between the raw data without ocular artifact and the cleaned data (CPD) before the earliest speech onset (700 ms) and the average of the speech onsets (900 ms), which indicates that the quality of the cleaned data is enough for subsequent EEG processing. The validation result shows a small difference during the pre-speech onset, which might also be caused by removing the ocular artifact.

\section{CONCLUSION}
\label{sec:conclusion}

We proposed a speech artifact removal method using tensor decomposition. Our proposed method surpassed the former methods (SAR-ICA and BSS-CCA) both in identifying speech artifacts (0.985) and producing cleaned data (0.101), validated by around 0.92-0.94. Both the evaluation and validation were calculated with the Pearson correlation coefficient using the grand-average of the related clusters and the lip EMG. Since the present study relied on visual inspection, future research should fully automate all of the steps needed for artifact identification.

\section{ACKNOWLEDGEMENT}
\label{sec:acknowledgement}

Part of this work was supported by JSPS KAKENHI (Grant Numbers JP17H06101, JP17K00237, and JP16K16172).

\bibliographystyle{IEEEtran}
\bibliography{references}

\nocite{*}

\end{document}